\documentclass[%
 reprint,
 amsmath,amssymb,
prl,
]{revtex4-1}

\usepackage{lipsum}
\usepackage{graphicx}
\usepackage{dcolumn}
\usepackage{bm}

\newcommand{\vecr}{\bm{r}}
\DeclareMathOperator{\sign}{sign}
\DeclareMathOperator{\E}{E}
\DeclareMathOperator{\Var}{Var}
\newcommand{\tot}[2]{{{#1}\to{#2}}}
\newcommand{\expct}[1]{\langle #1 \rangle}
\newcommand{\figref}[1]{Fig.~\ref{#1}}
\renewcommand{\eqref}[1]{Eq.~(\ref{#1})}
\newcommand{\pref}[1]{(\ref{#1})}
\renewcommand{\(}{\left(}
\renewcommand{\)}{\right)}
\renewcommand{\[}{\left[}
\renewcommand{\]}{\right]}

\begin{document}

\title{Lane formation and critical coarsening in a model of bacterial competition}

\author{Takuro Shimaya}
\email{t.shimaya@noneq.phys.s.u-tokyo.ac.jp}
\author{Kazumasa A. Takeuchi}
\email{kat@kaztake.org}
\affiliation{Department of Physics, Tokyo Institute of Technology, 2-12-1 Ookayama, Meguro-ku, Tokyo, 152-8551, Japan}%
\affiliation{Department of Physics, The University of Tokyo, 7-3-1 Hongo, Bunkyo-ku, Tokyo, 113-0033, Japan}%

\date{\today}

\begin{abstract}
We study competition of two non-motile bacterial strains in a three-dimensional channel numerically, and analyze how their configuration evolves in space and time.
We construct a lattice model that takes into account self-replication, mutation, and killing of bacteria.
When mutation is not significant, the two strains segregate
 and form stripe patterns along the channel.
The formed lanes are gradually rearranged, with increasing length scales
 in the two-dimensional cross-sectional plane.
We characterize it in terms of coarsening and phase ordering
 in statistical physics.
In particular, for the simple model without mutation and killing,
 we find logarithmically slow coarsening,
 which is characteristic of the two-dimensional voter model.
With mutation and killing, we find a phase transition from
 a monopolistic phase, in which lanes are formed and coarsened
 until the system is eventually dominated by one of the two strains,
 to an equally mixed and disordered phase without lane structure.
Critical behavior at the transition point is also studied and compared with
 the generalized voter class and the Ising class.
These results are accounted for by continuum equations,
 obtained by applying a mean field approximation along the channel axis.
Our findings indicate relevance of critical coarsening of \textit{two}-dimensional systems
 in the problem of bacterial competition within anisotropic \textit{three}-dimensional geometry.
\end{abstract}

\maketitle

\section{I.~Introduction}
Competition and evolution of multiple biological species,
 such as those in ecosystems, constitute one of the key situations
 where ideas of statistical physics can contribute
 to quantitative understanding of biological problems and vice versa
 \cite{Crow.Kimura-Book1970,Drossel-AP2001,Korolev2010,JSTATevo}.
Traditionally, theoretical approaches to such competition processes often
 assumed uniform systems without any spatial structure
 \cite{Crow.Kimura-Book1970,Turchin-Book2003},
 which correspond to studying well-mixed populations.
However, recent experiments have shown that,
 when multiple strains of bacteria are cultured on agar,
 nontrivial domain structures are formed,
 which then interplay with their population and evolutionary dynamics
 \cite{Hallatschek2007,Rudge.etal-ASB2013,Lloyd2015,Farrell2017,Mcnally2017}.
Formation of clonal domains, as well as their spatiotemporal evolution,
 were also observed in stem cell tissues,
 and shed light on mechanisms of homeostasis \cite{Klein2011,MESA2018677}.
Those studies have shown that many aspects of cell populations
 can be characterized by universal scaling laws
 developed in statistical physics,
 such as those for coarsening \cite{Mcnally2017,Klein2011,MESA2018677}
 and interface fluctuations \cite{Hallatschek2007,Farrell2017}.
Also backed by a surge of theoretical interests
 in evolutionary dynamics \cite{JSTATevo,Korolev2010}
 and active matter \cite{Marchetti.etal-RMP2013},
 interplay between competition and spatial degrees of freedom
 has aroused increasing attention.

Recent experimental developments on microfluidic devices
 \cite{VELVECASQUILLAS201028} add another aspect to this problem.
An advantage of microfluidic systems is that
 one has control over the system geometry.
Indeed, it is now clear that the system geometry can have a crucial impact
 on collective properties of cells \cite{Wu2017,Beppu2017}.
One of the common geometries for long time measurement is a channel
 with open ends
 used to characterize growth and division of single cells, cell lineage,
 statistical properties of cell populations, etc. \cite{Volfson2008,Mather2010,Wang.etal-CB2010,Boyer.etal-PB2011,Long2013,Hashimoto2016,Sheats170463}.
As such, it is also a natural geometry to use
 for studying competition problems.

Here we study competition of two non-motile bacterial strains
 in a channel with open ends.
The two strains are differently labeled but otherwise isogenic.
We devise a simple model to investigate
 the possible existence of universal macroscopic properties of the problem.
In its simplest version, the model consists of self-replication of cells,
 volume exclusion, and escape from the open ends.
Then we find that initially mixed populations spontaneously segregate,
 forming lane structures along the channel.
Spatiotemporal evolution of lanes can be characterized in terms of
 phase ordering in the cross section of the channel.
Remarkably, it turns out to show logarithmically slow coarsening,
 characteristic of the \textit{two}-dimensional voter model
 \cite{Scheucher1988,Dornic2001},
 although our model has \textit{three}-dimensional geometry.
We also generalize the model by introducing mutation and killing of bacteria,
 and find a transition from a monopolistic phase,
 in which lanes are formed and coarsened
 until the system is eventually dominated by one of the two strains,
 to an equally mixed and disordered phase without lane structure.
Moreover, the characteristics observed at the critical point
 between those two phases suggest the relevance
 of the generalized voter class \cite{Dornic2001,AlHammal2005}
 known from studies of so-called absorbing-state transitions
 \cite{Hinrichsen-AP2000},
 though the possibility of the Ising class is not rule out either.
These results are accounted for by continuum equations,
 which we obtain by the mean field approximation along the channel axis.


\section{II.~Model with self-replication only}
\begin{figure}[t]
\includegraphics{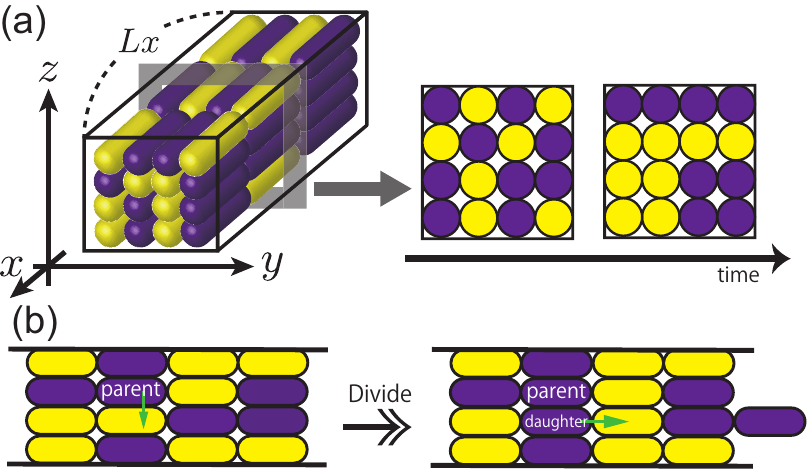}
\caption{
\label{fig-1} 
Illustration of the model with self-replication only. 
(a) Sketch of the system.
The channel is along the $x$-axis
 and filled with two strains of bacteria (yellow and purple).
If lanes are formed along the channel,
 their arrangement can be characterized by cross sections.
(b) Sketch of the time evolution rule (see text).
}
\end{figure}

We consider two strains of non-motile bacteria that self-replicate,
 inside an open channel with rectangular cross section (\figref{fig-1})
 as used in actual experiments \cite{Volfson2008,Mather2010,Hashimoto2016}.
The channel consists of a three-dimensional lattice
 of size $L_x\times{}L_y\times{}L_z$
 (see, e.g., \cite{Korolev2010,Mcnally2017,Thompson2011,Hashimoto2016}
 for the validity of lattice models to characterize
 statistical properties of cell populations).
The $x$-axis is taken along the channel.
We impose the open boundary condition at the channel ends
 and the periodic one at the walls.
Each site is occupied by a cell of genotype $s(x,y,z,t)\in\{-1,1\}$
 (shown in yellow and purple, respectively, in \figref{fig-1}).
Each cell has a division age $\tau_{\mathrm{rep}}$.
Following an experimental observation
 of \textit{Escherichia coli} \cite{Hashimoto2016}, 
 here we assume the gamma distribution for $\tau_{\mathrm{rep}}$
\footnote{
Iyer-Biswas \textit{et al.}
 \cite{IyerBiswas.etal-PRL2014,IyerBiswas.etal-PNAS2014}
 proposed the beta exponential distribution for the division time.
In practice, however, with the coefficient of variation
 (standard deviation-to-mean ratio) of their experimental data,
 and within the range of the histogram they showed,
 the gamma distribution has a similar shape and therefore can be regarded
 as an approximate distribution.
}.
When the division time comes, the cell replicates a daughter
 with the same genotype $s$ at one of the six nearest-neighbor sites.
This neighbor is chosen as follows:
 first the direction is chosen to be longitudinal or transverse,
 with respect to the channel axis,
 then one of the neighbors is selected at equal probability.
As a result, neighbors in the $x$ direction are chosen with probability $1/4$
 and those in the $yz$ direction with probability $1/8$.
This stochastic rule reflects the situation in which cells are partially
 oriented along the channel
 because of the excluded volume effect \cite{Volfson2008, Cho2007}.
Figure~\ref{fig-1}(b) illustrates an example in which the replication
 takes place in a direction perpendicular to the channel walls.
In this case, the generated daughter cell pushes the existing cell
 toward either end of the channel, which is again chosen randomly.
The row of cells is pushed thereby, and the one at the extremity is expelled
 from the system.
The total number of the cells is therefore conserved.
If the replication occurs along the channel,
 the row of cells is pushed similarly.

Both of the divided cells renew their $\tau_{\mathrm{rep}}$
 according to the gamma distribution. 
In the following, we fix the parameters of the gamma distribution
 so that the mean is $\E[\tau_{\mathrm{rep}}]=50$
 and the variance is $\Var[\tau_{\mathrm{rep}}]=200$.
Simulations were carried out
 by using Gillespie's algorithm \cite{Gillespie-JPC1977} with continuous time.

\begin{figure}[t]
\includegraphics[width=\hsize]{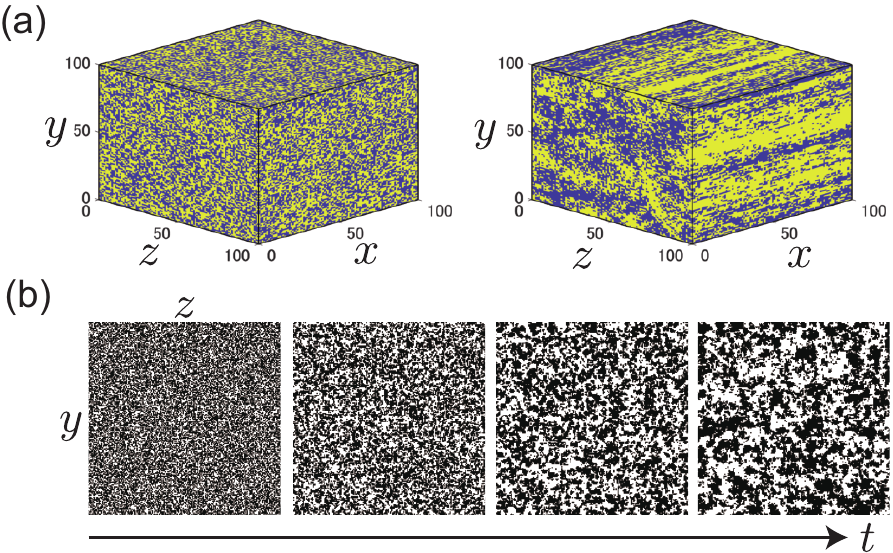}
\caption{
\label{fig-2}
Lane formation and subsequent coarsening
 in the model with self-replication only.
(a) Three-dimensional view of the system with $L_x=L_y=L_z=100$.
The two strains are indicated by yellow ($s=-1$)
 and purple ($s=+1$).
The left and right figures show the configuration
 at different times, $t=0$ (initial condition)
 and $t=50000$, respectively.
See also Movie~S1 \cite{SI}.
(b) Time evolution of the two-dimensional magnetization field $\phi(y,z,t)$
($\blacksquare$:$\phi>0$, $\square$:$\phi\leq0$)
 at $t=0,300,1500,5000$ from left to right.
The system size is $L_x=L_y=L_z=200$.
See also Movie~S2 \cite{SI}.
}
\end{figure}

Figure~\ref{fig-2}(a) and Movie~S1 \cite{SI} show
 time evolution of the system from a random initial condition.
The initial condition is generated by setting
 $s(x,y,z,0)=+1$ with probability $p_0$ and $s(x,y,z,0)=-1$ otherwise,
 independently at each site,
 with $p_0=0.5$ corresponding to the non-biased situation.
Then we find that the two, initially mixed strains of bacteria segregate
 in the course of time, forming lanes along the channel.
Moreover, typical width of those lanes grows with time (see the movie).
This suggests the relevance of coarsening
 and dynamic scaling in statistical physics \cite{Bray-AP1994},
 which describes, e.g., how the domains of up and down spins evolve
 in the ferromagnetic phase of the Ising model.
There is an obvious analogy because our variable $s$ is also dichotomous,
 but the time evolution of our model does not satisfy the detailed balance
 (in this sense non-equilibrium) and is anisotropic by construction.

\begin{figure*}[t]
    \centering
        \begin{tabular}{rl}
        \begin{minipage}{0.5\hsize}
            \includegraphics[width=\hsize]{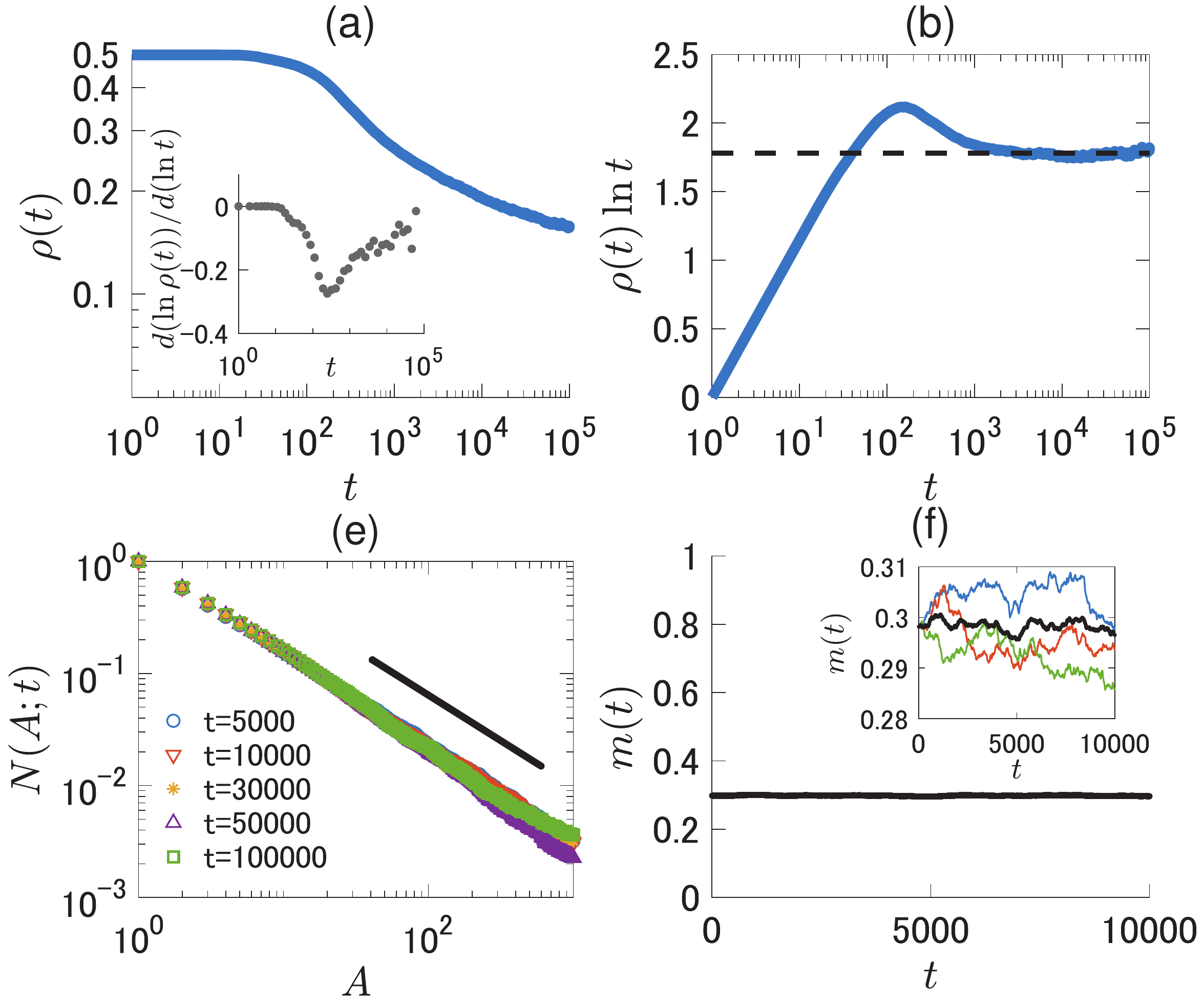}
        \end{minipage}
        \begin{minipage}{0.5\hsize}
            \includegraphics[width=\hsize]{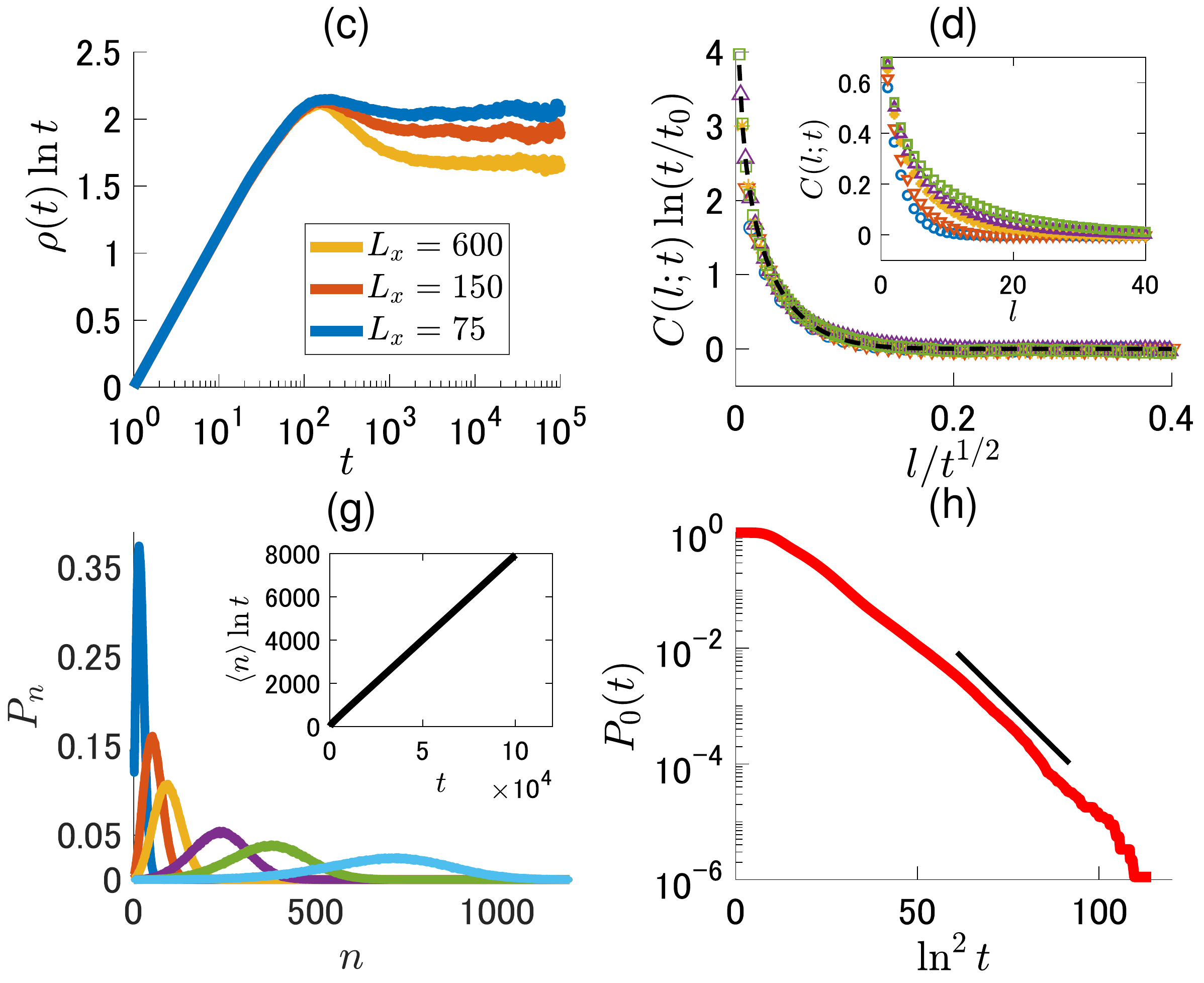}
        \end{minipage}
        \end{tabular}
\caption{
\label{fig-3}
Results for the model with self-replication only, averaged over 10 independent realizations.
The data in (a), (b) and (d)-(h) are obtained with the system size $L_x=L_y=L_z=300$.
The data in (a)-(e), (g) and (h) are obtained with unbiased initial conditions
 with $p_0=0.5$ ($m(0)\approx0$).
(a)(b) The interface density in the $\phi$-field, $\rho(t)$.
Raw data is shown in (a) in the log-log chart.
The inset of (a) shows the local exponent $d(\ln \rho(t))/d(\ln t)$.
The absence of a plateau region indicates that $\rho(t)$ does not decay by a power law.
The horizontal dashed line is a guide for the eyes,
 showing logarithmic decay of the data, $\rho(t)\sim{}1/\ln{}t$ (b).
The value of the local exponent tends to zero asymptotically (a, inset),
 being consistent with the logarithmic decay.
(c) Channel length ($L_x$) dependence of the interface density $\rho(t)$.
The size of the cross section is fixed at $L_y =L_z =150$.
Although the asymptotic value of $\rho(t) \ln t$ depends on $L_x$,
 the logarithmic decay of $\rho(t)$ is robust
 against changes in the channel aspect ratio.
(d) The spatial correlation function $C(l;t)$ at $t=5000(\circ),10000(\triangledown),30000(\ast),50000(\triangle),100000(\Box)$
 (from lower left to upper right in the inset). 
$t_0=300$ is used. 
The black dashed line shows the Ei function fitted to the data at $t=100000$.
(e) The cumulative distribution of domain size $A$, $N(A;t)$.
The solid line shows $N(A;t)\sim{}A^{-\tau}$ with $\tau=0.81(3)$.
(f) The total magnetization $m(t)$ for simulations
 from biased initial conditions with $p_0=0.65$.
The black bold line shows the ensemble average over 10 realizations,
 zoomed in the inset,
 and the thinner color lines in the inset show individual time series.
(g) The main panel shows the probability that $\sign[\phi(y,z,t)]$ changes $n$ times until time $t$, $P_n(t)$
 ($t=1000, 5000, 10000, 30000, 50000, 100000$, from upper left to lower right).
The inset shows how the average number of sign flips, $\langle n \rangle$ (averaged in space and over realizations), increases with $t$. Our observation is in agreement with the result for the voter model, $\left< n \right>\sim t/\ln t$ \cite{BenNaim1996}.
(h) Persistence probability $P_0(t)$, i.e., the probability that $\sign[\phi(y,z,t)]$ never changes until time $t$.
We find behavior consistent with the result for the voter model \cite{BenNaim1996}, 
$P_0(t)\sim \exp(-\mathrm{const.}\times \ln^2 t)$, indicated by the solid line.
}
\end{figure*}

To characterize the observed anisotropic coarsening,
 we introduce the following local ``magnetization''
\begin{equation}
\phi(y,z,t) := \frac{1}{L_x}\sum^{L_x}_x s(x,y,z,t),
\label{eq:phi}
\end{equation}
 which is a function of cross-sectional coordinates $(y,z)$ and time.
The sign of $\phi(y,z,t)$, denoted by $\sign[\phi(y,z,t)]$,
 indicates the strain that takes the majority
 in each line along the channel.
Figure~\ref{fig-2}(b) and Movie~S2 \cite{SI} show
 space-time evolution of $\sign[\phi(y,z,t)]$.
They clearly show the growth of length scales
 -- an important characteristic of coarsening processes --
 in cross sections.
On the other hand, the intricate structure of the observed patterns does
 not seem to be characterized by a single growing length scale;
 as a matter of fact, the domain interfaces are irregular down to the
 smallest length scale of the system, i.e., the lattice constant.
It is contrasted with coarsening in the ferromagnetic Ising model
 and that of other curvature-driven interfaces, for which
 interfaces are smoothed by effective surface tension \cite{Bray-AP1994}.

One of the standard method for characterizing coarsening
 is to measure the total length of the domain interfaces.
For our model, we use $\sign[\phi(y,z,t)]$ to determine the domains,
 and measure the interface density $\rho(t)$,
 defined by the fraction of site pairs with the opposite signs
 [\figref{fig-3}(a)(b)].
In contrast to usual curvature-driven coarsening,
 for which $\rho(t)$ typically decays by a power law \cite{Bray-AP1994},
 here $\rho(t)$ seems to decay more slowly [\figref{fig-3}(a)].
Indeed, if $\rho(t)\ln{}t$ is plotted instead [\figref{fig-3}(b)],
 we find an extended plateau asymptotically, which indicates
 $\rho(t)\sim{}1/\ln{}t$.
In fact, this logarithmic decay is known to be characteristic
 of the two-dimensional voter model \cite{Scheucher1988,Dornic2001},
 a simple model for opinion formation.
Similarity to the voter model is also apparent
 from the pattern evolution of $\sign[\phi(y,z,t)]$ [\figref{fig-2}(b)],
 which resembles that of the voter model \cite{Scheucher1988,Dornic2001}.
These results are robust against changes in the system aspect ratio,
 as we checked for both elongated $(L_x>L_y=L_z)$ and shortened $(L_x<L_y=L_z)$
 channels [\figref{fig-3}(c)].

The appearance of the characteristic coarsening of the voter model
 is further confirmed quantitatively.
For example, the largest length scale of the pattern is known to grow
 as $t^{1/2}$ in the voter model \cite{Cox1986,Scheucher1988,Dornic2001}.
A way to see this is to measure the spatial correlation function
\begin{equation}
 C(l;t) := \left< \phi(\bm{r}+\bm{l},t)\phi(\bm{r},t)\right> -\left< \phi(\bm{r},t) \right>^2,
\end{equation}
 with $\bm{r}:=(y,z)$ and $l:=|\bm{l}|$.
The tail of the measured correlation function is indeed more extended
 for larger times [\figref{fig-3}(d) inset],
 showing growth of the relevant length scale.
For the voter model, the asymptotic expression
 of the correlation function is known to be \cite{Krapivsky-PRA1992,Dornic1998}
\begin{equation}
 C(l;t) \simeq \frac{\mathrm{Ei}_1(l^2/2t)}{\ln(t/t_0)}, \label{eq:scorr}
\end{equation}
 with the exponential integral (Ei) function
 $\mathrm{Ei}_1(\xi):=\int^{\infty}_{\xi}w^{-1}e^{-w}\mathrm{d}w$
 and a microscopic time scale $t_0$, which is $1/16$ for the voter model
 but is in general a model-dependent quantity.
This form of rescaling is tested in \figref{fig-3}(d) main panel.
The data are found to overlap very well,
 being in remarkable agreement with the Ei function
 predicted for the voter model (dashed line).
We also measure the cumulative distribution of the domain area $A$ at time $t$,
 $N(A;t)$ [\figref{fig-3}(e)].
As opposed to the correlation function, the domain area distribution
 is governed by different length scales that coexist in the pattern,
 and as a result it is essentially independent of time.
We find a power-law distribution $N(A;t)\sim{}A^{-\tau}$,
 which implies fractal structure of the pattern.
We obtained an exponent value $\tau=0.81(3)$ from the data at $t=100000$,
 which is consistent with a past study on the voter model \cite{Scheucher1988}.
In addition, we also measure the total magnetization of the system,
 $m(t):=\expct{s(x,y,z,t)}$, and find that it remains statistically constant,
 even if we start from a biased initial condition [\figref{fig-3}(f)].
Statistical conservation of $m$
 is also an important characteristic of the voter model \cite{Dornic2001}.
Finally, agreement with the voter model is also seen
 in statistical properties of the change of $\sign[\phi(y,z,t)]$,
 such as the persistence probability $P_0(t)$
 and the average number of sign flips $\expct{n(t)}$ [\figref{fig-3}(g)(h)],
 studied numerically for the voter model in \cite{BenNaim1996}.


\begin{figure}[t]
\includegraphics{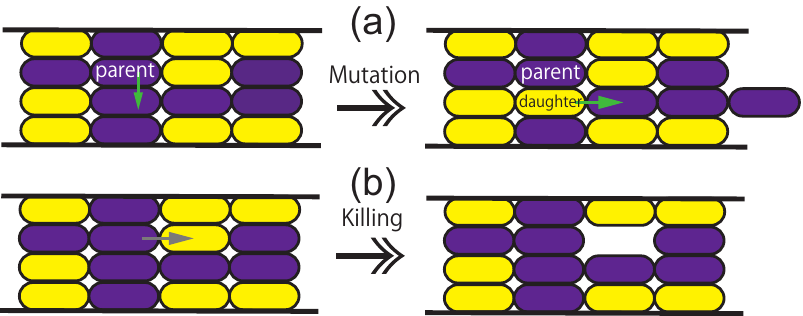}
\caption{
\label{fig-4} 
Mutation (a) and killing (b) considered in the extended model (see text).
}
\end{figure}

\section{III.~Model with mutation and killing}
To investigate the robustness of our results under more general situations,
 we extend our model by including mutation and killing of bacteria
 (\figref{fig-4}).
For simplicity, here the mutation is implemented by an event in which
 a parent cell generates a daughter with the genotype $s$,
 or allele, opposite to that of the parent [\figref{fig-4}(a)].
We assume that mutation occurs with probability $p_\mathrm{m}$
 at each replication.
We can also interpret this mutation as switching
  between bistable states of gene regulatory networks \cite{Sneppen2010}.
For the killing, we implement it by the following stochastic event,
 having in mind the bacterial type VI secretion system (T6SS)
 \cite{HO20149,Russell2014,Mcnally2017}.
When a cell decides to kill, it chooses a target randomly among the neighbors.
If and only if the chosen cell has the genotype different from the killer's,
 it is killed and a void is generated.
This void, encoded as $s=0$, can be taken by a cell generated at a later time.
A killing event occurs randomly and independently from replications.
The waiting time, $\tau_{\mathrm{kill}}$,
 is generated from the exponential distribution
 with mean $\E[\tau_\mathrm{kill}]$.
We define a parameter
 $C_\mathrm{k}:=\E[\tau_{\mathrm{rep}}]/\E[\tau_{\mathrm{kill}}]$.
The previous model without mutation and killing corresponds to taking
 $p_\mathrm{m}=0$ and $C_\mathrm{k}=0$.

\begin{figure}[t]
    \centering
        \begin{tabular}{cc}
        \begin{minipage}{\hsize}
            \includegraphics[width=\hsize]{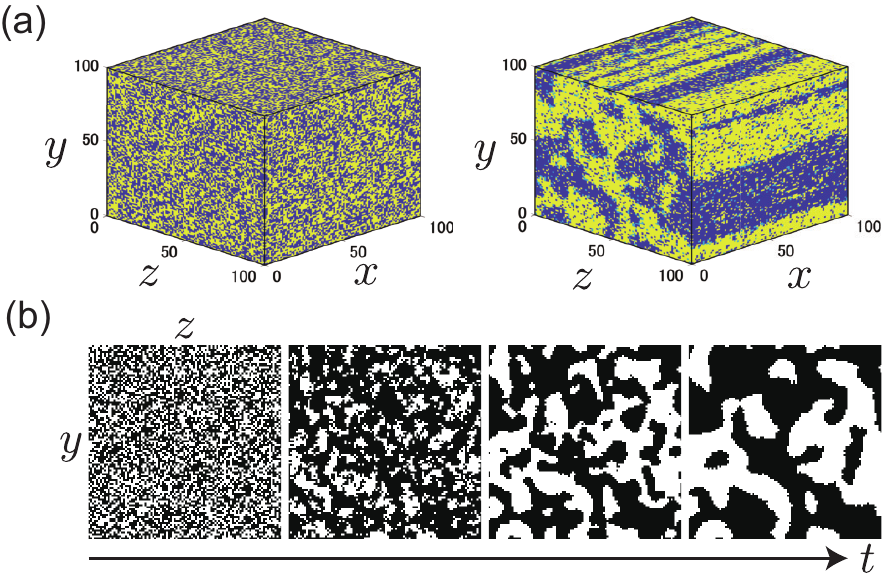}
        \end{minipage} \\
        \begin{minipage}{\hsize}
            \includegraphics[width=\hsize]{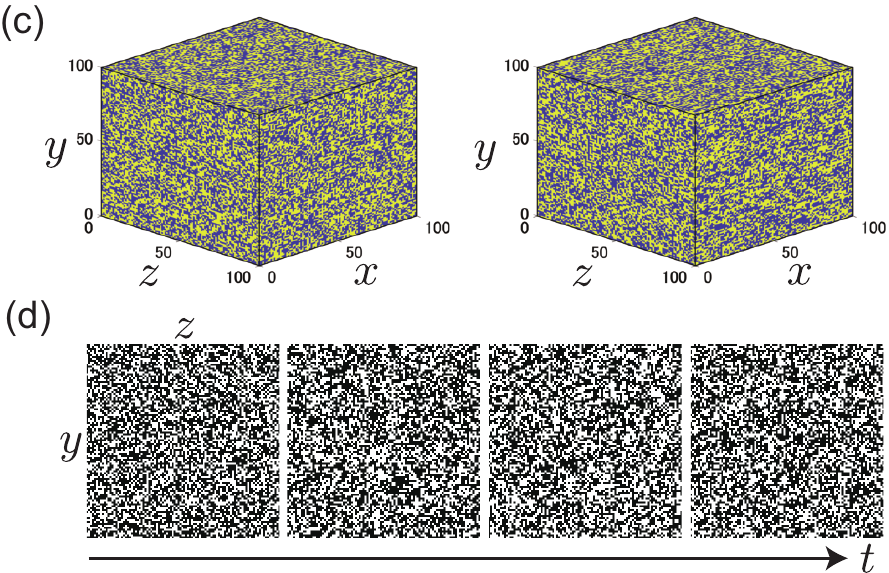}
        \end{minipage}
        \end{tabular}
\caption{
\label{fig-5}
Evolution of the model with mutation and killing.
The system size is $L_x=L_y=L_z=100$.
(a)(b) The monopolistic phase where killing processes are dominant
($p_\mathrm{m} = 0.3,\ C_\mathrm{k} = 2.0$).
(a) Three-dimensional view of the system.
There exist only few voids generated by killing (light blue).
The left and right figures show the configuration at different times, $t=0$ (initial condition) and $t=5000$, respectively.
See also Movie~S3 \cite{SI}.
(b) Time evolution of the two-dimensional magnetization field $\phi(y,z,t)$, at $t=0,500,1500,5000$ from left to right, for the realization shown in (a).
(c)(d) The data in the mixed phase where mutation is dominant
($p_\mathrm{m} = 0.3,\ C_\mathrm{k} = 0.05$) is shown in the same way. 
See also Movie~S4 \cite{SI}.
}
\end{figure}

\begin{figure}[t]
\includegraphics[scale=0.37]{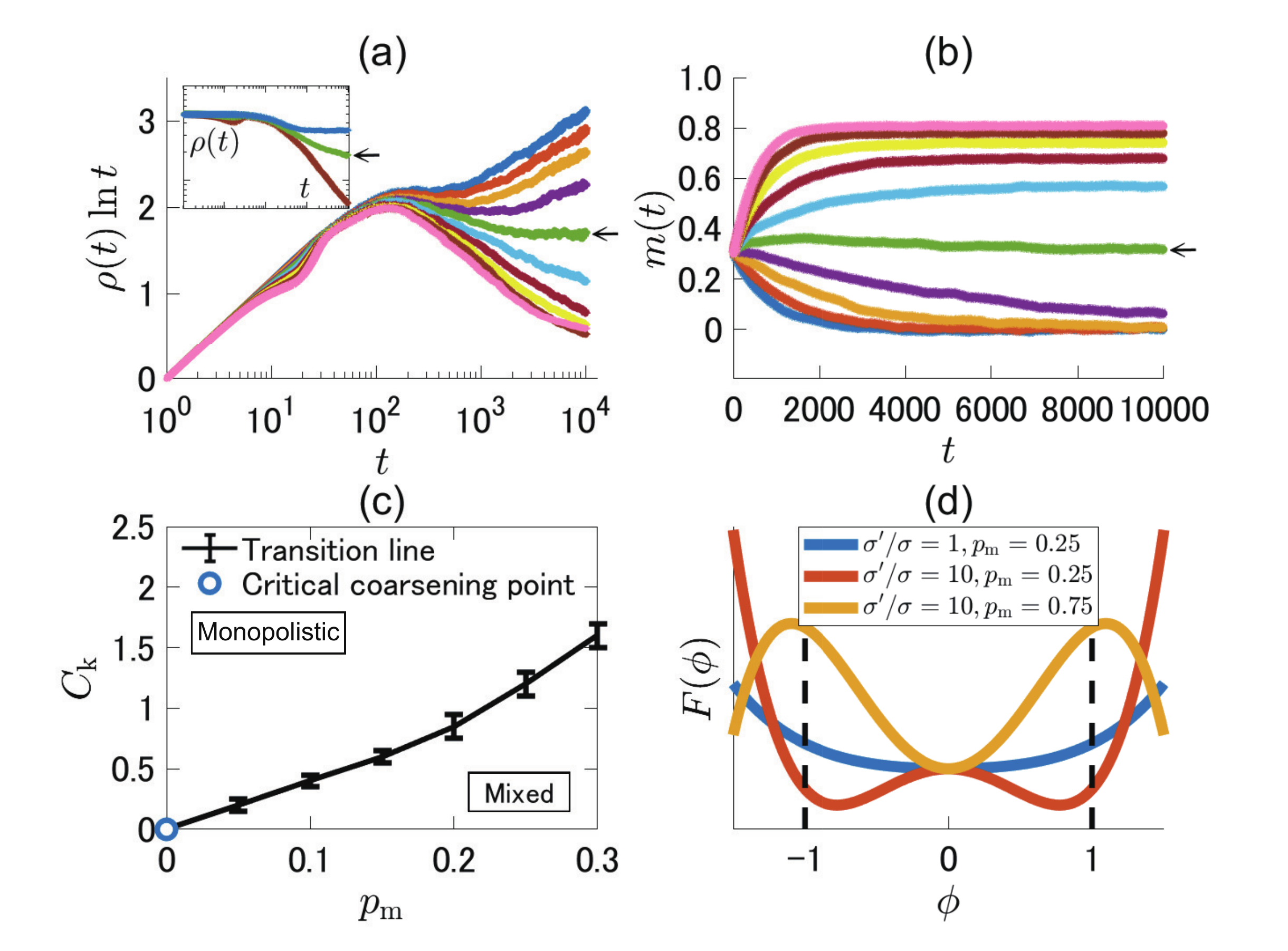}
\caption{
\label{fig-6} 
Results for the model with mutation and killing, with system size $L_x=100$ and $L_y=L_z=200$.
(a,b) The interface density $\rho(t)$ (a)
 and the total magnetization $m(t)$ (b) for different $C_\mathrm{k}$
 with $p_\mathrm{m}$ fixed at 0.05.
$C_\mathrm{k}$ is varied from 0 to 0.45
 [from top to bottom for (a), from bottom to top for (b)].
The initial conditions are $m(0)\approx0$ in (a) and
 $m(0)\approx0.3$ in (b), respectively.
Each data set was taken from a single realization.
The green curve indicated by the arrow ($C_\mathrm{k}=0.2$)
 is considered to be closest to the transition point.
(c) Phase diagram in the $(p_\mathrm{m},C_\mathrm{k})$ plane.
(d) Profile of the free energy density $F(\phi)$ [\eqref{eq:freee}]
 in the obtained continuum equations.
}
\end{figure}

Carrying out simulations
 for various values of $p_\mathrm{m}$ and $C_\mathrm{k}$,
 we find that lane formation and subsequent coarsening occur as well,
 for relatively small $p_\mathrm{m}$ or large $C_\mathrm{k}$.
Figure~\ref{fig-6}(a) inset and \figref{fig-6}(b) show
 the density of interfaces in $\sign[\phi(y,z,t)]$, $\rho(t)$,
 and the total magnetization $m(t)$, respectively, for $p_\mathrm{m}=0.05$
 and $C_\mathrm{k}$ varied from 0 to 0.45.
For large $C_\mathrm{k}$, we observe lane formation and coarsening
 (\figref{fig-5}(a)(b) and Movie~S3 \cite{SI}),
 accompanied by decrease of $\rho(t)$ [\figref{fig-6}(a) inset].
However, unlike the voter-type coarsening in the previous model,
 the interfaces are smoother [\figref{fig-5}(a)(b)],
 $\rho(t)$ decreases faster than $1/\ln{}t$
 [\figref{fig-6}(a) main panel, lower curves], and $m(t)$ is not conserved
 but takes a non-zero asymptotic value determined by the choice
 of the parameter values [\figref{fig-6}(b) upper curves].
This means that the system is eventually dominated by one of the two strains.
In contrast, if $C_\mathrm{k}$ is small, lanes are not formed
 (\figref{fig-5}(c)(d) and Movie~S4),
 $\rho(t)$ stops decreasing [\figref{fig-6}(a) inset top curve],
 and $m(t)$ vanishes [\figref{fig-6}(b) lower curves].
In other words, the two strains remain mixed and equally populated.
Figure~\ref{fig-6}(c) shows a phase diagram
 in the $(p_\mathrm{m},C_\mathrm{k})$ plane,
 where the monopolistic (ordered) and mixed (disordered) phases
 are bordered by a transition line.
Near the transition,
 our data seem to indicate $\rho(t)\sim{}1/\ln{}t$ and constant $m(t)$
 [\figref{fig-6}(a,b), green curves indicated by the arrows].
The simpler case without mutation and killing,
 $(p_\mathrm{m},C_\mathrm{k})=(0,0)$, corresponds to the endpoint
 of the transition line [\figref{fig-6}(c)].

These results can be interpreted as follows.
First of all, while mutation obviously makes the configuration more disordered,
 killing actually plays a role analogous
 to the Ising ferromagnetic interaction.
This is because, firstly,
 killing occurs only between cells of different genotypes,
 and secondly, the void left by the killed cell
 is eventually taken by a daughter from one of the neighbors.
As a result of those competing effects of mutation and killing,
 the ordered and disordered phases appear,
 similarly to the ferromagnetic Ising model.
Moreover, the presence of the Ising-like ferromagnetic interaction
 also implies that the interfaces are now endowed
 with effective surface tension \cite{Bray-AP1994},
 which can explain why those in the ordered phase are smoother
 than the voter-type coarsening observed at the transition.

Concerning the transition, in the literature
 it is known that the characteristic coarsening of the voter model
 represents a broad class of phase transitions into absorbing states
 \cite{Hinrichsen-AP2000} in the presence of the Ising-like up/down symmetry,
 called the generalized voter universality class
 \cite{Dornic2001,AlHammal2005}.
Systems in the generalized voter class usually have two symmetric absorbing states,
 labeled by ``spin'' variable $+1$ and $-1$.
The defining feature of those absorbing states is that bulk nucleation
 of the opposite spin is forbidden; in other words,
 once the spin variables become globally $+1$ or $-1$,
 this uniform configuration is kept forever.
Such systems can show two different phase transitions,
 one for spontaneous symmetry breaking of magnetization,
 and the other for whether the system eventually falls into
 one of the two absorbing states.
According to the established scenario \cite{AlHammal2005},
 if these transitions occur separately, the former is in the Ising class
 and the latter is in the directed percolation class.
However, the two transitions can also occur simultaneously in generic models,
 and in this case the voter universality class arises.
Now, let us recall that our model has the $Z_2$ symmetry
 (symmetry with respect to $s \leftrightarrow -s$)
 and that the characteristics of the two-dimensional voter model
 were clearly identified at $(p_\mathrm{m}, C_\mathrm{k}) = (0,0)$,
 i.e., the endpoint of the transition line.
Therefore, it is reasonable to expect that the transition in the general case
 is described by either the generalized voter class or the Ising class
 in two dimensions (see also Al Hammal et al.'s theory for transitions
 in the presence of two symmetric absorbing states \cite{AlHammal2005}).
These possibilities are tested in the following.

\begin{figure*}[t]
    \centering
        \begin{tabular}{cc}
        \begin{minipage}{0.9\hsize}
            \includegraphics[width=\hsize]{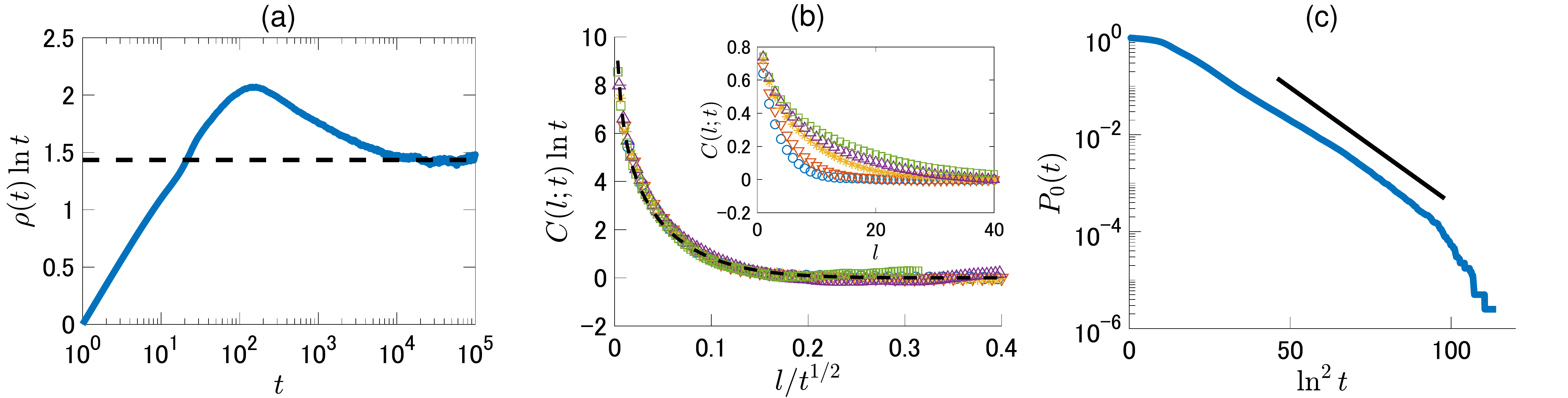}
        \end{minipage} \\
        \begin{minipage}{0.9\hsize}
            \includegraphics[width=\hsize]{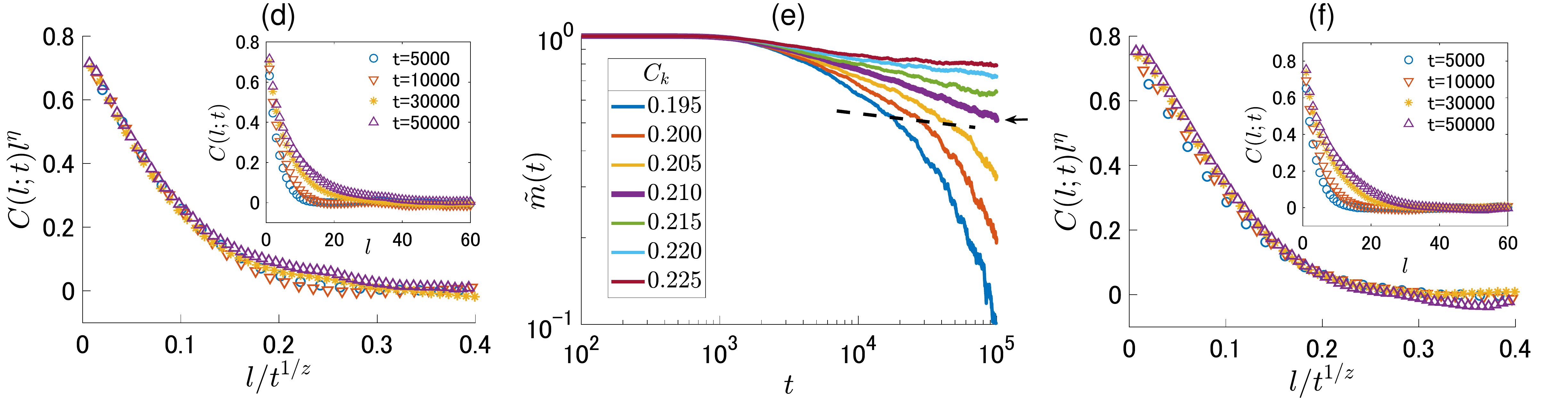}
        \end{minipage}
        \end{tabular}
\caption{
\label{fig-7}
Statistical properties near the critical point for the model with mutation and killing.
The system size is $L_x=100$ and $L_y=L_z=200$.
Non-biased random initial conditions are used ($p_0=0.5, m(0)\approx0$).
(a)(b)(c) Test of the generalized voter class with mutation and killing near the transition
($p_\mathrm{m} = 0.05,\ C_\mathrm{k} = 0.215$).
(a) The interface density $\rho(t)$ showing logarithmic decay
$\rho(t)\sim{}1/\ln{}t$ at late times.
(b) The spatial correlation function $C(l;t)$ at $t=5000(\circ),10000(\triangledown),30000(\ast),50000(\triangle),100000(\Box)$.
The black dashed line shows the Ei function fitted to the data at $t=100000$ [see \eqref{eq:scorr}].
The inset shows law data without rescaling.
(c) Persistence probability $P_0(t)$.
The solid line indicates the result for the voter model
\cite{BenNaim1996}, $P_0(t)\sim \exp(-\mathrm{const.}\times \ln^2 t)$.
(d)(e)(f) Test of the Ising critical behavior.
(d) The spatial correlation function $C(l;t)$
for $p_\mathrm{m}=0.05$ and $C_\mathrm{k}=0.21$,
at which the Ising scaling law \pref{IsingCorr} seems to be
satisfied most accurately for $p_\mathrm{m}$ fixed at 0.05
and varied $C_\mathrm{k}$.
Here the known values of the critical exponents
for the two-dimensional Ising class, $\eta=1/4$ and $z=2.183$ \cite{Dammann1993}, were used.
The inset shows the raw data.
(e) Decay of the cross-sectional magnetization $\tilde{m}(t)$,
for different $C_\mathrm{k}$ with $p_\mathrm{m}$ fixed at 0.05.
The completely dominated initial condition is used: $p_0=1$ ($m(0)=1$).
The purple curve indicated by the arrow corresponds to the putative critical point $(p_\mathrm{m}, C_\mathrm{k}) = (0.05,0.21)$ obtained in panel (d),
showing a power law decay $\tilde{m}(t) \sim t^{-\alpha}$ with
$\alpha = 0.170(20)$.
It is significantly different from the critical magnetization decay
for the two-dimensional Ising class, $\alpha = \beta/\nu z \simeq 0.057$
(dashed line).
On the other hand, the data for $C_\mathrm{k}=0.22$ give $\alpha=0.056(9)$, which coincides with the Ising value within the range of error.
(f) The spatial correlation function for $p_\mathrm{m}=0.05$ and $C_\mathrm{k}=0.22$.
The other conditions are the same as in (d).
}
\end{figure*}

First, we test the possibility of the two-dimensional generalized voter class.
In the presence of mutation, although strict absorbing states do not exist,
our data near the transition seem to indicate
a set of characteristic properties of the voter model,
specifically, $\rho(t)\sim{}1/\ln{}t$ and constant $m(t)$
[\figref{fig-6}(a,b), green curves indicated by the arrows],
suggesting the relevance of the generalized voter class.
With fixed $p_\mathrm{m} (=0.05)$ and varying $C_\mathrm{k}$, 
 we further determine the critical point by detecting the logarithmic decay
 of the interface density $\rho(t)$.
Simulations for different values of $C_\mathrm{k}$, with $L_x=100, L_y=L_z=200$
 and non-biased random initial conditions ($p_0=0.5$),
 indicated that $C_\mathrm{k} = 0.215$ was closest to the critical point
 [\figref{fig-7}(a)].
With this set of the parameter values,
 we evaluate the spatial correlation function $C(l;t)$,
 as well as the persistent probability $P_0(t)$ that
 $\sign[\phi(y,z,t)]$ never changes until time $t$ [\figref{fig-7}(b)(c)].
We find that both data seem to agree
 with the predicted behavior of the voter model:
 the spatial correlation function is described by \eqref{eq:scorr}
 and the persistent probability is consistent
 with the predicted behavior for the voter model
 \cite{BenNaim1996}, $P_0(t)\sim \exp(-\mathrm{const.}\times \ln^2 t)$.
It is possible that, after the lane formation,
 the presence of the majority strain in each line along the channel
 may play the role of a (nearly) absorbing state,
 since it takes long time for the majority to change
 if the channel length is long enough.

Next, we consider the possibility of the two-dimensional Ising class.
 To determine the critical point, we consider that the magnetization
  may not be the optimal quantity to use, because it decays algebraically
  both in the ordered phase and at the critical point.
 Therefore, we use instead the spatial correlation function, which is known
  to satisfy the following scaling form at the critical point:
 \begin{equation}
     C(l;t)=l^{-\eta}f_c(l;t^{1/z})
     \label{IsingCorr},
 \end{equation}
  with critical exponents $\eta$ and $z$ \cite{Onuki, Dammann1993}.
 We then obtain a set of parameter values
  $(p_\mathrm{m}, C_\mathrm{k}) = (0.05, 0.21)$
  that gave the best collapse for the scaling form \eqref{IsingCorr}
  [\figref{fig-7}(d)],
  using simulations with non-biased random initial conditions
  ($p_0=0.5, m(0)\approx0$).
 At this putative critical point, we measure decay of the magnetization
  in the cross-sectional plane, $\tilde{m}(t)$,
  defined by the mean value of the sign of $\phi(y,z,t)$:
 \begin{equation}
     \tilde{m}(t) = \frac{1}{L_yL_z}\sum_{y,z} \sign[\phi(y,z,t)]
     \label{tildem}.
 \end{equation}
 Starting simulations from a single genotype, i.e., $p_0 = 1$ and $m(0) = 1$,
  we find a power-law decay $\tilde{m}(t)\sim 1/t^\alpha$
  with $\alpha=0.170(20)$ [\figref{fig-7}(e)].
 In the case of the two-dimensional Ising class,
  the magnetization indeed decays by a power law
 \begin{equation}
     \tilde{m}(t)\sim 1/t^{\frac{\beta}{\nu z}}
     \label{IsingMag},
 \end{equation}
  with critical exponents $\beta, \nu, z$ \cite{STAUFFER1997344},
  but the value of the decay exponent is $\beta/\nu z\simeq0.057$
  and far from the one obtained from the simulations, $\alpha=0.170(20)$.
 
As an alternative approach, we may also use \eqref{IsingMag} to determine
 the critical point.
This gave $C_\mathrm{k}=0.22$ for $p_\mathrm{m}=0.05$,
 for which the exponent $\alpha$ was estimated at $\alpha = 0.056(9)$,
 close enough to the Ising value $\beta/\nu z\simeq0.057$.
At this set of the parameter values,
 we tested the scaling form \eqref{IsingCorr} for the correlation function
 [\figref{fig-7}(f)].
Although the results apparently showed systematic deviation
 from the collapse at small $l$, we consider that
 careful finite-size analysis with larger system sizes is needed to draw
 a firm conclusion.
Therefore, with the present data sets,
 while our results are consistent with the generalized voter class
 in a number of statistical properties, the possibility of the Ising class
 cannot be ruled out either.
Conclusive determination of the universality class
 is an important problem left for future studies.

\section{IV.~Theory}
The numerical results presented so far can be understood
 by means of continuum equations, which we obtain in the following
 by a mean-field-like approximation.
The variable to use is the local magnetization field $\phi(\vecr,t)$
 with $\vecr=(y,z)$ in the cross-sectional plane.
Suppose, at position $\vecr_i$, $\phi_i:=\phi(\vecr_i,t)$
 changes by $\Delta\phi_i$ during a small time step $\Delta{}t$.
$\Delta\phi_i$ can be expressed as follows,
\begin{eqnarray}
\Delta\phi_i = \Delta\phi_{i\to i} + \sum_{j\in\{\mathrm{n.n\ of\ i}\}}\Delta\phi_{j\to i} \label{eq:sumphi},
\end{eqnarray}
 where n.n. refers to the nearest neighbors in the $yz$ plane,
 and $\Delta\phi_\tot{j}{i}$ denotes the contribution
 from the line $\bm{r}_j$ to $\bm{r}_i$.
The change $\Delta\phi_\tot{j}{i}$ results from
 replications, mutations, and killing events
 that occur locally and independently
 at all pairs of sites between the two lines
 (or along the single line if $i=j$).
Therefore, by the central limit theorem, it can be approximated by
\begin{eqnarray}
\Delta\phi_{j \to i} = \E[\Delta\phi_{j\to i}] + \sqrt{\Var[\Delta\phi_{j\to i}]} \epsilon_{j\to i}(t) \label{eq:meanf}, 
\end{eqnarray}
where $\epsilon_\tot{j}{i}(t)$ is white Gaussian noise with $\langle\epsilon_\tot{j}{i}\rangle=0$ and $\langle\epsilon_\tot{j}{i}(t)\epsilon_\tot{j'}{i'}(t')\rangle=\delta_{ii'}\delta_{jj'}\delta_{tt'}$.

The mean $\E[\Delta\phi_\tot{j}{i}]$
 and the variance $\Var[\Delta\phi_\tot{j}{i}]$ can be evaluated
 by considering, for each type of events,
 the Poisson distribution for the number of the events
 and the probability that such an event changes the magnetization $\phi_i$,
 within the mean-field approximation (see Appendix for details).
For simplicity, here we consider that replications and killing attempts occur
 at constant rates $\sigma$ and $\sigma'$, respectively
 (roughly $\sigma\approx1/\E[\tau_\mathrm{rep}]$
 and $\sigma'\approx1/\E[\tau_\mathrm{kill}]$),
 and the void generated by killing is filled immediately
 by replication from a neighboring site.
Then, taking the limit $\Delta{}t\to0$ and coarse-graining in space,
 we obtain
\begin{align}
 \frac{\partial \phi(\vecr,t)}{\partial t}
 = &-\frac{\delta F(\phi)}{\delta \phi} + A_1\nabla^2\phi \notag \\
 &\qquad + \sqrt{A_2(1-\phi^2) + A_3 \phi^2} ~\eta(\vecr,t)
 \label{eq:pfeq}
\end{align}
 with a Landau-like free energy density
\begin{equation}
 F(\phi) = A_4 \phi^2 + A_5 \phi^4  \label{eq:freee}.
\end{equation}
 and coefficients
\begin{align}
    &A_1 = a^2\[ \frac{\sigma}{8}(1-p_\mathrm{m}) + \frac{\sigma'}{16}\],
    && \notag \\
    &A_2 = \frac{a^2}{L_x}(\sigma+2\sigma'),
    &&A_3 = \frac{2a^2}{L_x}\sigma p_\mathrm{m}, \notag \\
    &A_4 = -\left( \frac{1-2p_\mathrm{m}}{4}\sigma' - p_\mathrm{m}\sigma \right),
    &&A_5 = \frac{1-2p_\mathrm{m}}{8}\sigma'.  \label{eq:coeff}
\end{align}
Here, $\eta(\vecr,t)$ is white Gaussian noise
 with $\langle\eta(\vecr,t)\rangle=0$ and 
 $\langle\eta(\vecr,t)\eta(\vecr',t')\rangle=\delta(\vecr'-\vecr)\delta(t-t')$,
 and $a$ is the lattice constant.

\begin{figure}[t]
    \includegraphics[width=\hsize]{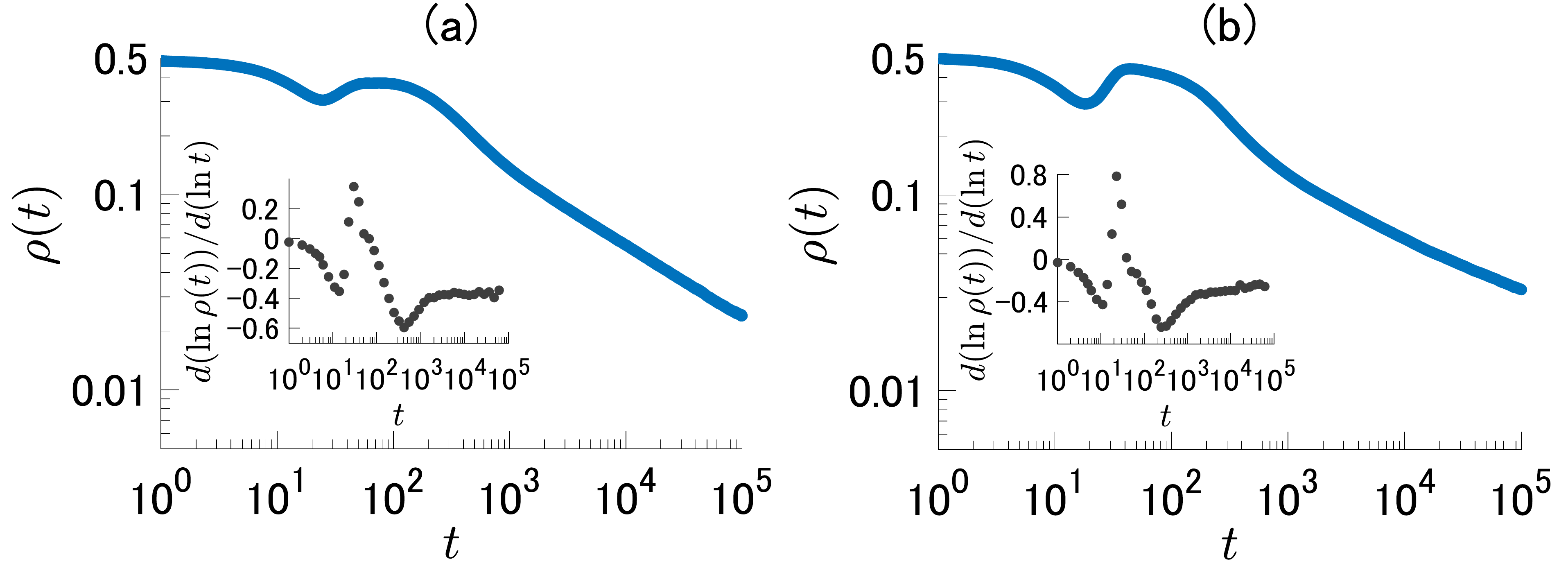}
    \caption{
    \label{fig-8} 
    Decay of the interface density $\rho(t)$ in the presence of killing but without mutation (specifically, $p_\mathrm{m}=0, C_\mathrm{k}=1$, deep in the monopolistic phase).
    10 independent realizations were used.
    Non-bised initial conditions are used.
    The system size is $L_x=20,\ L_y=L_z=200$ for (a)
     and $L_x=100, L_y=L_z=200$ for (b).
    The insets show the local exponent $d(\ln \rho(t))/d(\ln t)$.
    Although the data suggest power-law decay instead of the logarithmic one, the exponent takes values significantly smaller
     (in the absolute value) than that of the Ising model with Glauber dynamics,
     $-0.5$.
    The exponent value becomes even smaller for the longer channel.
    }
\end{figure}

Several remarks are now in order.
First, in the case without mutation and killing ($p_\mathrm{m}=\sigma'=0$), 
 we have $A_1,A_2>0$ and $A_3=A_4=A_5=0$.
Then \eqref{eq:pfeq} becomes the Lengevin description of the voter model
 \cite{Dickman1995,Munoz1997}, which underpins our observation
 of the voter-type coarsening in this case (\figref{fig-3}).
Second, though both coefficients of $F(\phi)$
 can change the sign in general [Fig.\ref{fig-6}(d)],
 for $p_\mathrm{m}<1/2$, $A_5$ remains positive, while $A_4$ changes the sign
 at $p_m=\frac{\sigma'}{4\sigma+2\sigma'}\approx\frac{C_\mathrm{k}}{4+2C_\mathrm{k}}$.
This underlies the transition observed in Fig.\ref{fig-6}(c).
Finally, since $-\frac{\delta{}F(\phi)}{\delta\phi}=\[\frac{1-2p_\mathrm{m}}{2}\sigma'(1-\phi^2)-2p_\mathrm{m}\sigma\]\phi$,
 in the absence of mutation ($p_\mathrm{m}=0$), 
 the completely monopolistic situations $\phi(\vecr,t)=\pm1$
 correspond to the two absorbing states of \eqref{eq:pfeq}.
Further, \eqref{eq:pfeq} in this case takes the form
 of the continuum equation proposed by Al Hammal {\it et al.}
 for the generalized voter universality class \cite{AlHammal2005}.
If $p_\mathrm{m}\neq0$, $\phi(\vecr,t)=\pm1$ are not absorbing any more,
 but our numerical data near the transition seem to remain consistent with
 the generalized voter class, though the Ising class is not ruled out either
 as we already discussed.
In contrast, in the monopolistic (ordered) phase,
 the ordering process seems to be driven by curvature,
 or effective surface tension between the two domains [\figref{fig-5}(a)(b)].
However, while theoretically $\rho(t)\sim{}t^{-1/2}$ is expected in this case,
 in our simulations $\rho(t)$ decays significantly more slowly [\figref{fig-8}(a)(b)].
This apparent discrepancy, which may be due to an approximation
 made to derive the continuum equation, needs to be elucidated.

\section{V.~Summary}
In this work, we constructed a model of competition between two strains
 of non-motile bacteria in a channel.
In the simplest situation driven only by self-replications,
 we numerically found that the two strains segregate
 and form lane structures along the channel.
The lanes then gradually thicken.
This process was characterized in the cross-sectional plane
 and turned out to be governed
 by the coarsening process of the two-dimensional voter model.
In the presence of killing and mutation,
 we revealed a transition between the mixed phase and the monopolistic phase.
In the mixed phase, lane formation does not occur
 and the two strains are well-mixed, with equal proportion on average.
In contrast, in the monopolistic case, one of the two strains dominates,
 though a coarsening process qualitatively similar
 to that of the Ising ferromagnet.
Near the transition, our data seem to be consistent
 with the generalized voter class,
 which includes the self-replication-only case
 at the end of the transition line,
 though the possible realization of the Ising class in the presence of mutation
 is not ruled out either.

We should note that,
 at the price of access to large-scale statistical properties,
 our model assumes an idealized situation,
 which in many aspects oversimplifies actual bacterial competition
 and ecosystems.
However, statistical properties of the voter model identified in this work
 are known to be universal in the absence of long-range interactions
 in a space of interest (cross-sectional plane in our context).
Indeed, if cells are aligned as we consider in the model,
 previous studies suggested that such long-range interactions
 were absent or very weak \cite{Volfson2008,Mather2010, Cho2007}.
One can therefore expect the same behavior to arise 
 even in the presence of complex biochemical and mechanical interactions.
Of course it is of crucial importance to test this prediction
 in real experiments and/or more realistic models.
Finally, while we considered two neutral strains in this work,
 competition between two non-neutral strains is also of considerable interest.
According to our theory, this problem is expected to be described
 by an asymmetric free energy density,
 and in the monopolistic phase,
 one of the two local minima becomes a metastable state.
It is then interesting to test, e.g.,
 the relevance of transient dynamics around a metastable state
 \cite{Nikolay2003,Fiasconaro2005}
 in the context of population and evolutionary dynamics.

\begin{acknowledgments}
\section{Acknowledgments}
We acknowledge discussions with R. A. L. Almeida, H. Chat\'e, I. Dornic
 and N. Mitarai.
We thank Y. T. Fukai for having drawn our attention to \cite{Mcnally2017},
 and T. P. Shimizu for a help in program coding.
This work is supported in part by KAKENHI
 from Japan Society for the Promotion of Science
 (No. JP25103004, JP16H04033, JP16K13846) and
 by the grants associated with ``Planting Seeds for Research'' Program
 and Suematsu Award of Tokyo Tech.
\end{acknowledgments}

\appendix*
\section{APPENDIX: DERIVATION OF THE CONTINUUM EQUATION}
Here we derive the continuum equation \pref{eq:pfeq},
 which describes the time evolution of the local magnetization $\phi(\vecr,t)$
 with corss-sectional coordinates $\vecr = (y,z)$ at coarse-grained scales.
We start from Eqs.~\pref{eq:sumphi} and \pref{eq:meanf} for the lattice model:
\begin{eqnarray}
\Delta \phi_i &=& \Delta \phi_{i \to i} + \sum_{j\in \{ \mathrm{n.n.\ of\ }i\}} \Delta \phi_{j \to i}, \label{eq:incre} \\
\Delta \phi_{j \to i} &=& \E[\Delta \phi_{j \to i}] + \sqrt{\Var[\Delta \phi_{j \to i}]}\epsilon_{j\to i}(t), \label{eq:aa}
\end{eqnarray}
 where $\E[\cdot]$ and $\Var[\cdot]$ denote the mean and the variance,
 respectively,
 and $\epsilon_{j\to i}(t)$ is white Gaussian noise with $\left< \epsilon_{j\to i}(t)\right>=0$ and  $\left< \epsilon_{j\to i}(t)\epsilon_{j'\to i'}(t')\right>=\delta_{ii'}\delta_{jj'}\delta_{tt'}$.
$\Delta \phi_{j \to i}$ is the variation of $\phi_i:=\phi(\vecr_i,t)$
 due to stochastic events that occur in a neighboring line $\vecr_j$,
 during a small time step $\Delta t$.
Such a variation occurs, for example,
 when a cell in the line $\vecr_j$ replicates,
 produces its daughter in the line $\vecr_i$,
 and this repels a cell of the other strain (opposite spin)
 at either channel end.
Similarly, $\phi_i$ varies when a cell in the line $\vecr_j$ kills a cell
 (of the different genotype) in the line $\vecr_i$, and this void is filled
 by replication of a neighboring cell,
 which is assumed here to occur immediately for the sake of simplicity.
Such series of events can occur only when the pair (or the triplet) of sites
 have appropriate combinations of genotype $s$.
The probability of having such combinations,
 $P^{\pm,\mathrm{rep}}_{j\to i}$ and $P^{\pm,\mathrm{kill}}_{j\to i}$,
 for replication and killing processes, respectively,
 with the double sign indicating whether $\phi_i$ increases or decreases,
 can be expressed as functions of $\phi_i$ and $\phi_j$
 by employing a mean-field approximation.
With those probabilities,
 as well as the number of replication events $\lambda_{\mathrm{rep}}(x,y,z)$
 and that of killing events $\lambda_{\mathrm{kill}}(x,y,z)$
 at a site $(x,y,z)$ (with $\vecr_i = (y,z)$)
 during the time step $\Delta t$,
 $\Delta \phi_{j \to i}$ can be expressed as
\begin{widetext}
\begin{eqnarray}
    \Delta \phi_{j \to i} = && \left(+\frac{2}{L_x} \right) \left\{ \sum^{L_xP^{+,\mathrm{rep}}_{j\to i}}_x\lambda_{\mathrm{rep}}(x,y,z)+ \sum^{L_xP^{+,\mathrm{kill}}_{j\to i}}_x\lambda_{\mathrm{kill}}(x,y,z)\right\} \nonumber \\
    && \qquad + \left( -\frac{2}{L_x} \right)\left\{ \sum^{L_xP^{-,\mathrm{rep}}_{j\to i}}_x\lambda_{\mathrm{rep}}(x,y,z) + \sum^{L_xP^{-,\mathrm{kill}}_{j\to i}}_x\lambda_{\mathrm{kill}}(x,y,z) \right\}\label{eq:aaa}.    
\end{eqnarray}
\end{widetext}
Here, $L_x$ is the channel length,
 or the total number of the cells in each lane.

Now, for simplicity, we assume that replication and killing events
 occur independently at constant rates
 $\sigma$ ($\approx 1/\E[\tau_{\mathrm{rep}}]$)
 and $\sigma'$ ($\approx 1/\E[\tau_{\mathrm{kill}}]$), respectively.
Then the number of such events obey the Poisson distribution, so that
 $\E[\lambda_{\mathrm{rep}}(x,y,z)] = \Var[\lambda_{\mathrm{rep}}(x,y,z)] = \sigma\Delta t$ and $\E[\lambda_{\mathrm{kill}}(x,y,z)] = \Var[\lambda_{\mathrm{kill}}(x,y,z)] = \sigma'\Delta t$.
We thereby obtain
\begin{widetext}
\begin{eqnarray}
    \E[ \Delta \phi_{j \to i} ] = && \left( +2\right) \( \sigma\Delta tP^{+,\mathrm{rep}}_{j\to i} + \sigma'\Delta t P^{+,\mathrm{kill}}_{j\to i}\) +  \left( -2\right) \(\sigma\Delta t P^{-,\mathrm{rep}}_{j\to i}  +  \sigma'\Delta tP^{-,\mathrm{kill}}_{j\to i}\), \label{eq:a} \\
    \Var[ \Delta \phi_{j \to i} ] = && \left( +\frac{2}{L_x} \right)^2 \left( \sigma\Delta t L_xP^{+,\mathrm{rep}}_{j\to i}+\sigma'\Delta t L_x P^{+,\mathrm{kill}}_{j\to i} \right)  + \left( -\frac{2}{L_x} \right)^2 \left( \sigma\Delta t L_xP^{-,\mathrm{rep}}_{j\to i}+\sigma'\Delta t L_xP^{-,\mathrm{kill}}_{j\to i} \right). \label{eq:b}
\end{eqnarray}
\end{widetext}

The probabilities
 $P^{\pm,\mathrm{rep}}_{j\to i}$ and $P^{\pm,\mathrm{kill}}_{j\to i}$
 are evaluated by applying a mean-field approximation
 along each line of the channel.
For the replication, with the effect of mutation taken into account,
 we obtain
\begin{align}
P^{\pm,\mathrm{rep}}_{j\to i} &=& \frac{1}{8}(1 - p_\mathrm{m})\frac{1 \pm \phi_j}{2}\frac{1 \mp \phi_i}{2} + \frac{1}{8}p_\mathrm{m} \frac{1 \mp \phi_j}{2}\frac{1 \mp \phi_i}{2} \nonumber \\
&& \qquad \qquad (j \neq i) \label{eq:c},  \\
P^{\pm,\mathrm{rep}}_{i\to i} &=& \frac{1}{2}(1 - p_\mathrm{m})\frac{1 \pm \phi_j}{2}\frac{1 \mp \phi_i}{2} + \frac{1}{2}p_\mathrm{m} \frac{1 \mp \phi_j}{2}\frac{1 \mp \phi_i}{2} \label{eq:d},
\end{align}
 where $p_\mathrm{m}$ is the probability that mutation occurs
 at each replication.
Here, the factor $\frac{1 \pm \phi_j}{2}$ corresponds to the probability
 that the cell to replicate in the line $\vecr_j$ is the strain $s=\pm 1$,
 and $\frac{1 \pm \phi_i}{2}$ to the probability that the cell to be repelled
 from the channel in the line $\vecr_i$ is the strain $\pm 1$.
The coefficient $1/8$ in \eqref{eq:c} is the probability that
 the specific line $\vecr_j ~(\neq \vecr_i)$ is chosen
 as the position of the daughter cell.
It is simply replaced with $1/2$ for in-line replications.
Similarly, for killing processes, we obtain
\begin{eqnarray}
P^{\pm,\mathrm{kill}}_{j\to i} &=& \frac{1 \pm \phi_j}{2}\frac{1}{8}\frac{1 \mp \phi_i}{2}G^\pm_i \qquad (j \neq i), \nonumber \\
P^{\pm,\mathrm{kill}}_{i\to i} &=& \frac{1 \pm \phi_i}{2}\frac{1}{2}\frac{1 \mp \phi_i}{2}G^\pm_i,
\end{eqnarray}
 where $G^\pm_i$ is the probability that a cell of strain $\pm 1$
 self-replicates to fill the void generated by killing.
It is given by
\begin{eqnarray}
G^\pm_i = && (1-p_\mathrm{m})\left( \frac{1}{2}\frac{1 \pm \phi_i}{2} + \sum_{j' \in \{\mathrm{n.n.\ of}\ i\}} \frac{1}{8}\frac{1 \pm \phi_{j'}}{2} \right) \nonumber \\
&& + p_\mathrm{m}\left( \frac{1}{2}\frac{1 \mp \phi_i}{2} + \sum_{j' \in \{\mathrm{n.n.\ of}\ i\}} \frac{1}{8}\frac{1 \mp \phi_{j'}}{2} \right) \label{eq:o},
\end{eqnarray}
 where n.n. refers to the nearest neighbors in the $yz$ plane.

\subsection{A.~The case without mutation and killing (self-replication only)}
Let us first consider the simplest case without mutation and killing
 ($p_\mathrm{m}=0$ and $\sigma'=0$), in which we found
 characteristic coarsening of the two-dimensional voter model.
From \eqref{eq:a}-\pref{eq:d}, we obtain
\begin{eqnarray}
\E[ \Delta \phi_{j \to i} ] &=& \frac{1}{8}(\phi_j-\phi_i) \sigma\Delta t, \nonumber \\
\Var[ \Delta \phi_{j \to i}] &=& \frac{1}{4}(1-\phi_j\phi_i)\left( \frac{\sigma\Delta t}{L_x} \right) \qquad (j \neq i), \label{eq:h}
\end{eqnarray}
 and
\begin{eqnarray}
\E[ \Delta \phi_{i \to i} ] &=& 0, \nonumber \\
\Var[ \Delta \phi_{i \to i} ] &=& (1-\phi_i^2)\left( \frac{\sigma\Delta t}{L_x} \right). \label{eq:f}
\end{eqnarray}
Therefore, by \eqref{eq:incre} and \pref{eq:aa}, we have
\begin{widetext}
\begin{equation}
\Delta\phi_i = \frac{1}{8}\sigma\Delta t \sum_{j\in \{\mathrm{n.n.\ of}\ i \}}(\phi_j-\phi_i)+\sum_{j\in \{\mathrm{n.n.\ of}\ i \}}\sqrt{\frac{\sigma\Delta t}{4L_x}(1-\phi_j \phi_i)}\epsilon_{j\to i}(t)+\sqrt{\frac{\sigma\Delta t}{L_x}(1-\phi_i^2)}\epsilon_{i\to i}(t) \label{eq:k}.
\end{equation}
This can be rewritten as
\begin{eqnarray}
\frac{\Delta\phi_i}{\Delta t} = \frac{\sigma}{8}\sum_{j\in \{\mathrm{n.n.\ of}\ i \}}(\phi_j-\phi_i)+\sqrt{\frac{\sigma}{4L_x\Delta t}\left\{ \sum_{j\in \{\mathrm{n.n.\ of}\ i \}}(1-\phi_j \phi_i) + 4(1-\phi_i^2)\right\} }\epsilon_{i}(t) \label{eq:kk},
\end{eqnarray}
\end{widetext}
where $\epsilon_{i}(t)$ is white Gaussian noise with
 $\left< \epsilon_{i}(t)\right>=0$ and
 $\left< \epsilon_{i}(t)\epsilon_{i'}(t')\right>=\delta_{ii'}\delta_{tt'}$.
Note that the first term of the right-hand side of \eqref{eq:kk}
 is a discrete Laplacian.

Now we coarse-grain the description,
 by replacing the discrete coordinates $\vecr_i$ with continuous ones $\vecr$
 and differences with derivatives,
 and take the limit $\Delta t \to 0$.
With the lattice constant $a$, we obtain
\begin{eqnarray}
\partial_t\phi(\vecr,t) = \frac{a^2 \sigma}{8}\nabla^2\phi + \sqrt{\frac{2a^2\sigma}{L_x}(1-\phi^2)}\eta(\vecr,t), \label{eq:l}
\end{eqnarray}
 with $\eta(\vecr,t)$ white Gaussian noise in continuous space and time,
 which satisfies $\left< \eta(\vecr,t) \right>=0$ and $\left< \eta(\vecr,t)\eta(\vecr',t') \right>=\delta(\vecr'-\vecr)\delta(t-t')$.
Here we used the relationship
 $\eta(\vecr,t) \simeq \epsilon_i(t)/\sqrt{a^2 \Delta t}$ that ensures
\begin{eqnarray}
\int \mathrm{d}\vecr \int \mathrm{d}t \left<\eta(\vecr,t)\eta(\bm{r'},t') \right> = 1. \label{eq:ll}
\end{eqnarray}
Importantly, the obtained equation \pref{eq:l}
 is exactly the Langevin description of the voter model,
 proposed by earlier studies \cite{Dickman1995, Munoz1997}.
This underpins our numerical observation of the voter-type coarsening
 presented in Fig.~\ref{fig-3}.

\subsection{B.~The general case with mutation and killing}
For the general case with arbitrary $p_\mathrm{m}$ and $\sigma'$,
 we obtain
\begin{widetext}
\begin{eqnarray}
\E[ \Delta \phi_{j \to i} ] &=& \frac{\sigma\Delta t}{8}(1-p_\mathrm{m})(\phi_j-\phi_i) + \frac{\sigma'\Delta t}{16}(\phi_j-\phi_i)  \nonumber \\
&& \qquad +\frac{\sigma'\Delta t}{32}(1-2p_\mathrm{m})(1-\phi_j\phi_i)\left( \phi_i +\sum_{j'\in \{\mathrm{n.n.\ of}\ i \}}\frac{\phi_{j'}}{4} \right) - \frac{\sigma\Delta t}{8}p_\mathrm{m}(\phi_j+\phi_i) \qquad (j \neq i), \label{eq:r} \\
\Var[\Delta \phi_{j \to i}] &=& \frac{1}{4}\frac{\sigma\Delta t}{L_x}(1-p_\mathrm{m})(1-\phi_j\phi_i) + \frac{1}{8}\frac{\sigma'\Delta t}{L_x}(1-\phi_j\phi_i) + \frac{1}{4}\frac{\sigma\Delta t}{L_x}p_\mathrm{m}(1+\phi_j\phi_i) \nonumber \\
&&\qquad +\frac{1}{16}\frac{\sigma' \Delta t}{L_x}(1-2p_m)(\phi_j-\phi_i)\left( \phi_i + \sum_{j'\in \{\mathrm{n.n.\ of}\ i \}}\frac{\phi_{j'}}{4} \right)\qquad (j \neq i), \label{eq:s}
\end{eqnarray}
 and
\begin{eqnarray}
\E[ \Delta \phi_{i \to i} ] &=& \frac{\sigma'\Delta t}{8}(1-2p_\mathrm{m})(1-\phi_i^2)\left( \phi_i + \sum_{j'\in \{\mathrm{n.n.\ of}\ i \}}\frac{\phi_{j'}}{4} \right)-\sigma\Delta t p_\mathrm{m}\phi_i \label{eq:p}, \\
\Var[ \Delta \phi_{i \to i}] &=& \frac{\sigma\Delta t}{L_x}(1-p_\mathrm{m})(1-\phi_i^2) + \frac{1}{2}\frac{\sigma'\Delta t}{L_x}(1-\phi_i^2) + \frac{\sigma\Delta t}{L_x}p_\mathrm{m}(1+\phi_i^2) \label{eq:q}. \\
\end{eqnarray}
Combining \eqref{eq:incre} and \pref{eq:aa}, we obtain
\begin{align}
\frac{\Delta \phi_i}{\Delta t} &= \frac{\sigma'\Delta t}{8}(1-2p_\mathrm{m})\left(1-\phi_i^2+\sum_{j\in \{\mathrm{n.n.\ of}\ i \}}\frac{1-\phi_{j}\phi_i}{4}\right)\left( \phi_i +\sum_{j\in \{\mathrm{n.n.\ of}\ i \}}\frac{\phi_{j}}{4} \right) -\sigma p_m\left(\phi_i +\sum_{j\in \{\mathrm{n.n.\ of}\ i \}}\frac{\phi_{j}+\phi_i}{8}\right) \notag \\
 & \qquad + \left\{ \frac{\sigma}{8}(1-p_\mathrm{m})+\frac{\sigma'}{16} \right\}\sum_{j\in \{\mathrm{n.n.\ of}\ i \}}(\phi_j-\phi_i) + \sqrt{\Var[\Delta\phi_\tot{i}{i}] + \sum_{j \in \{\mathrm{n.n.\ of}\ i\}} \Var[\Delta\phi_\tot{j}{i}]} \epsilon_i(t).
\end{align}
\end{widetext}
Then, carrying out the same coarse-graining and the continuous-time limit
 as in the previous section, we finally arrive at \eqref{eq:pfeq}:
\begin{eqnarray}
\partial_t \phi(\vecr,t) &=& - \frac{\delta F(\phi)}{\delta \phi} + a^2\left[ \frac{\sigma}{8}(1-p_\mathrm{m})+\frac{\sigma'}{16} \right]\nabla^2\phi   \nonumber \\
&& \qquad + \sqrt{\frac{a^2}{L_x}\left[ (2\sigma+\sigma')(1-\phi^2)+2\sigma p_\mathrm{m} \phi^2\right]}\eta(\vecr,t) \nonumber \\ \label{eq:pfeq}
\end{eqnarray}
 with
\begin{eqnarray}
F(\phi) = - \left( \frac{1-2p_\mathrm{m}}{4}\sigma' - p_\mathrm{m}\sigma \right)\phi^2 + \frac{1-2p_\mathrm{m}}{8}\sigma'\phi^4 \label{eq:freee}.\ \ 
\end{eqnarray}

\section{SUPPLEMENTAL MOVIE DESCRIPTIONS}
\noindent
{\bf Movie S1:}\\
Time evolution of the model with self-replication only.
The left surface is the channel outlet, while the top and right surfaces are the boundaries.
The two strains are indicated by yellow and purple.
The system size is $L_x = L_y = L_z = 100$. See also Fig.~\ref{fig-2}.\\
{\bf Movie S2:}\\
Time evolution of the two-dimensional magnetization field $\phi(y,z,t)$ ($\blacksquare$:$\phi>0$, $\square$:$\phi\leq0$),
 for the model with self-replication only.
 The system size is $L_x = L_y = L_z = 200$.\\
{\bf Movie S3:}\\
Time evolution of the model with mutation and killing, in the monopolistic phase ($p_\mathrm{m} = 0.3,\ C_\mathrm{k} = 2.0$).
The left surface is the channel outlet, while the top and right surfaces are the boundaries.
The two strains are indicated by yellow and purple.
The system size is $L_x = L_y = L_z = 100$. See also \figref{fig-5}(a)(b).\\
{\bf Movie S4:}\\
Time evolution of the model with mutation and killing,
 in the mixed phase ($p_\mathrm{m} = 0.3,\ C_\mathrm{k} = 0.05$).
The left surface is the channel outlet, while the top and right surfaces are the boundaries.
The two strains are indicated by yellow and purple.
The system size is $L_x = L_y = L_z = 100$. See also \figref{fig-5}(c)(d).\\

\bibliography{./ccbc-8}

\end{document}